\documentclass[a4paper,conference]{IEEEtran}
\usepackage{graphicx,ifthen}
\usepackage[cmex10]{amsmath}
\usepackage{psfrag,amssymb}
\usepackage{mathtools}
\usepackage{url}

\usepackage[noadjust]{cite}
\usepackage{color,pst-plot,stfloats,pst-sigsys}
\usepackage{pstricks-add}
\usepackage{subfigure}
\usepackage{ifthen}

\newtheorem{thm}{Theorem}
\newtheorem{lem}{Lemma}
\newtheorem{cor}{Corollary}
\newtheorem{prop}{Proposition}

\newtheorem{remark}{Remark}

\newtheorem{definition}{Definition}

\title{On the Information Dimension\\ of Multivariate Gaussian Processes}

\author{
\IEEEauthorblockN{
  Bernhard C. Geiger\IEEEauthorrefmark{1}%
, Tobias Koch\IEEEauthorrefmark{2}\IEEEauthorrefmark{3}
}
\IEEEauthorblockA{
\IEEEauthorrefmark{1}%
Signal Processing and Speech Communication Laboratory, Graz University of Technology, Graz, Austria\\
\IEEEauthorrefmark{2}%
Signal Theory and Communications Department, Universidad Carlos III de Madrid, 28911, Legan\'es, Spain\\
\IEEEauthorrefmark{3}%
 Gregorio Mara\~n\'on Health Research Institute, 28007, Madrid, Spain.\\
Emails: geiger@ieee.org, koch@tsc.uc3m.es}
}

\newcommand{\pdf}[1]{f_{#1}}
\newcommand{\pdfg}[1]{g_{#1}}
\newcommand{\psdt}[1]{S_{#1}(\theta)}

\newcommand{\sdft}[1]{F_{#1}(\theta)}
\newcommand{\sdf}[1]{F_{#1}}
\newcommand{\dsdft}[1]{F'_{#1}(\theta)}
\newcommand{\dsdf}[1]{F'_{#1}}
\newcommand{\cov}[1]{K_{#1}}
\newcommand{\covt}[1]{K_{#1}(\tau)}

\newcommand{\ent}[1]{H\left(#1\right)}
\newcommand{\diffent}[1]{h(#1)}
\newcommand{\mutinf}[1]{I(#1)}
\newcommand{\kld}[2]{D(#1\|#2)}

\newcommand{\entrate}[1]{\bar{H}\left(#1\right)}

\newcommand{\infodim}[1]{d(#1)}

\newcommand{\dimpinrate}[1]{d^*(#1)}

\newcommand{\dom}[1]{\mathcal{#1}}

\newcommand{\lebesgue}{\lambda}

\newcommand{\expec}[1]{\mathbb{E}\left(#1\right)}

\renewcommand{\det}{\mathrm{det}}
\newcommand{\e}[1]{\mathrm{e}^{#1}}

\newcommand{\covmat}[1]{C_{#1}}

\newcommand{\limk}{\lim_{k\to\infty}}
\newcommand{\limm}{\lim_{m\to\infty}}
\newcommand{\qRV}[2]{[#1]_#2}
\newcommand{\diff}{\mathrm{d}}

\newcommand{\Xb}{\boldsymbol{X}}

\newcommand{\Yb}{\boldsymbol{Y}}
\newcommand{\Nb}{\boldsymbol{N}}
\newcommand{\Wb}{\boldsymbol{W}}
\newcommand{\Zb}{\boldsymbol{Z}}
\newcommand{\Ub}{\boldsymbol{U}}

\renewcommand{\dimpinrate}{d(\{\Xb_t\})}
\renewcommand{\qRV}[2]{[#1]_{#2}}
\newcommand{\qRVm}[1]{\qRV{#1}{m}}

\newcommand{\trans}[1]{#1^{\textnormal{\textsf{\tiny T}}}} 
\newcommand{\bfmu}{{\boldsymbol{\mu}}}

\newcommand{\pcovt}[1]{\overline{\cov{#1}}(\tau)}

\renewcommand{\e}[1]{e^{#1}}
\renewcommand{\ent}[1]{H\left(#1\right)}
\renewcommand{\entrate}[1]{H'\left(#1\right)}
\renewcommand{\diffent}[1]{h\left(#1\right)}

\renewcommand{\liminf}{\varliminf}
\renewcommand{\limsup}{\varlimsup}

\renewcommand{\expec}[1]{\mathsf{E}\left[#1\right]}

\newcommand{\mat}[1]{#1}

\begin{document}

\maketitle

\begin{abstract}
The authors have recently defined the R\'enyi information dimension rate $d(\{X_t\})$ of a stationary stochastic process $\{X_t,\,t\in\mathbb{Z}\}$ as the entropy rate of the uniformly-quantized process divided by minus the logarithm of the quantizer step size $1/m$ in the limit as $m\to\infty$ (B. Geiger and T. Koch, ``On the information dimension rate of stochastic processes," in \emph{Proc.\ IEEE Int.\ Symp.\ Inf.\ Theory (ISIT)}, Aachen, Germany, June 2017). For Gaussian processes with a given spectral distribution function $F_X$, they showed that the information dimension rate equals the Lebesgue measure of the set of harmonics where the derivative of $F_X$ is positive. This paper extends this result to multivariate Gaussian processes with a given matrix-valued spectral distribution function $F_{\mathbf{X}}$. It is demonstrated that the information dimension rate equals the average rank of the derivative of $F_{\mathbf{X}}$. As a side result, it is shown that the scale and translation invariance of information dimension carries over from random variables to stochastic processes.

\renewcommand{\thefootnote}{}
\footnote{The work of Bernhard C. Geiger has been funded by the Erwin Schr\"odinger Fellowship J 3765 of the Austrian Science Fund. The work of Tobias Koch has received funding from the European Research Council (ERC) under the European Union's Horizon 2020 research and innovation programme (grant agreement number 714161), from the Spanish Ministerio de Econom\'ia y Competitividad under Grants TEC2013-41718-R, RYC-2014-16332 and TEC2016-78434-C3-3-R (AEI/FEDER, EU), and from the Comunidad de Madrid under Grant S2103/ICE-2845.}
\end{abstract}
\setcounter{footnote}{0}

\section{Introduction}\label{sec:intro}

In 1959, R\'enyi \cite{Renyi_InfoDim} proposed the information dimension and the $d$-dimensional entropy to measure the information content of general random variables (RVs). In recent years, it was shown that the information dimension is of relevance in various areas of information theory, including rate-distortion theory, almost lossless analog compression, or the analysis of interference channels. For example, Kawabata and Dembo \cite{Kawabata_RDDim} showed that the information dimension of a RV is equal to its rate-distortion dimension, defined as twice the rate-distortion function $R(D)$ divided by $-\log(D)$ in the limit as $D\downarrow 0$. Koch \cite{Koch_SLB} demonstrated that the rate-distortion function of a source with infinite information dimension is infinite, and that for any source with finite information dimension and finite differential entropy the Shannon lower bound on the rate-distortion function is asymptotically tight. Wu and Verd\'u \cite{Wu_Renyi} analyzed both linear encoding and Lipschitz decoding of discrete-time, independent and identically distributed (i.i.d.), stochastic processes and showed that the information dimension plays a fundamental role in achievability and converse results. Wu \emph{et al.} \cite{Wu_IF} showed that the degrees of freedom of the $K$-user Gaussian interference channel can be characterized through the sum of information dimensions. Stotz and B\"olcskei \cite{Stotz_IF} later generalized this result to vector interference channels. 

In~\cite{GeigerKoch_ISIT,GeigerKoch_DimRate}, we proposed the information dimension rate as a generalization of information dimension from RVs to univariate (real-valued) stochastic processes. Specifically, consider the stationary process $\{X_t,t\in\mathbb{Z}\}$, and let $\{\qRVm{X_t},\,t\in\mathbb{Z}\}$ be the process obtained by uniformly quantizing $\{X_t\}$ with step size $1/m$. We defined the information dimension rate $d(\{X_t\})$ of $\{X_t\}$ as the entropy rate of $\{\qRVm{X_t}\}$ divided by $\log m$ in the limit as $m\to\infty$ \cite[Def.~2]{GeigerKoch_DimRate}. We then showed that, for any stochastic process, $d(\{X_t\})$ coincides with the rate-distortion dimension of $\{X_t\}$~\cite[Th.~5]{GeigerKoch_DimRate}. We further showed that for stationary Gaussian processes with spectral distribution function $\sdf{X}$, the information dimension rate $d(\{X_t\})$ equals the Lebesgue measure of the set of harmonics on $[-1/2,1/2]$ where the derivative of $\sdf{X}$ is positive~\cite[Th.~7]{GeigerKoch_DimRate}. This implies an intuitively appealing connection between the information dimension rate of a stochastic process and its bandwidth.

In this work, we generalize our definition of $d(\{X_t\})$ to multivariate processes. Consider the $L$-variate (real-valued) stationary process $\{\Xb_t\}$, and let $\{\qRVm{\Xb_t}\}$ be the process obtained by quantizing every component process of $\{\Xb_t\}$ uniformly with step size $1/m$. As in the univariate case, the information dimension rate $d(\{\Xb_t\})$ of $\{\Xb_t\}$ is defined as the entropy rate of $\{\qRVm{\Xb_t}\}$ divided by $\log m$ in the limit as $m\to\infty$. Our main result is an evaluation of $d(\{\Xb_t\})$ for $L$-variate Gaussian processes with (matrix-valued) spectral distribution function $\sdf{\Xb}$. We demonstrate that in this case $d(\{\Xb_t\})$ equals the Lebesgue integral of the rank of the derivative of $\sdf{\Xb}$. As a corollary, we show that the information dimension rate of univariate complex-valued Gaussian processes is maximized if the process is proper, in which case it is equal to twice the Lebesue measure of the set of harmonics where the derivative of its spectral distribution function $\sdf{X}$ is positive.

As side results, we show that $d(\{\Xb_t\})$ is scale and translation invariant. These properties are known for the information dimension of RVs (cf.~\cite[Lemma~3]{Wu_PhD}), but they do not directly carry over to our definition of $d(\{\Xb_t\})$, which is why we state them explicitly in this paper.

\section{Notation and Preliminaries}\label{sec:prelim}

\subsection{Notation}
We denote by $\mathbb{R}$, $\mathbb{C}$, and $\mathbb{Z}$ the set of real numbers, the set of complex numbers, and the set of integers, respectively. We use a calligraphic font, such as $\dom{F}$, to denote other sets, and we denote complements as $\mathcal{F}^\mathsf{c}$.

The real and imaginary part of a complex number $z$ are denoted as $\mathfrak{Re}(z)$ and $\mathfrak{Im}(z)$, respectively, i.e., $z=\mathfrak{Re}(z)+\imath\mathfrak{Im}(z)$ where $\imath=\sqrt{-1}$. The complex conjugate of $z$ is denoted as $z^*$.

We use upper case letters to denote deterministic matrices and boldface lower case letters to denote deterministic vectors. The transpose of a vector or matrix is denoted by $\trans{(\cdot)}$; the Hermitian transpose by $
(\cdot)^\mathsf{H}$. The determinant of a matrix $\mat{A}$ is $\det \mat{A}$.

We denote RVs by upper case letters, e.g., $X$. For a finite or countably infinite collection of RVs we abbreviate $X_{\ell}^k\triangleq (X_{\ell},\dots,X_{k-1},X_k)$, $X_\ell^\infty \triangleq (X_{\ell},X_{\ell+1},\dots)$, and $X_{-\infty}^k \triangleq (\dots,X_{k-1},X_k)$. Random vectors are denoted by boldface upper case letters, e.g., $\Xb\triangleq\trans{(X_1,\dots,X_L)}$. Univariate discete-time stochastic processes are denoted as $\{X_t,\,t\in\mathbb{Z}\}$ or, in short, as $\{X_t\}$. For $L$-variate stochastic processes we use the same notation but with $X_t$ replaced by $\Xb_t\triangleq(X_{1,t},\dots,X_{L,t})$. We call $\{X_{i,t},\,t\in\mathbb{Z}\}$ a \emph{component process}. 

We define the quantization of $X$ with precision $m$ as
\begin{equation}\label{eq:quantization}
 \qRV{X}{m} \triangleq \frac{\lfloor mX\rfloor}{m}
\end{equation}
where $\lfloor a\rfloor$ is the largest integer less than or equal to $a$. Likewise, $\lceil a \rceil$ denotes the smallest integer greater than or equal to $a$. We denote by $\qRV{X_{\ell}^k}{m} = (\qRV{X_{\ell}}{m},\ldots,\qRV{X_k}{m})$ the component-wise quantization of $X_{\ell}^k$ (and similarly for other finite or countably infinite collections of RVs and random vectors). Likewise, for complex RVs $Z$ with real part $R$ and imaginary part $I$, the quantization $\qRV{Z}{m}$ is equal to $\qRV{R}{m}+\imath \qRV{I}{m}$. We define $\dom{C}(z_1^k,a)\triangleq[z_1,z_1+a) \times\cdots\times[z_k,z_k+a)$. Thus, $\dom{C}(z_1^k,a)$ is a $k$-dimensional hypercube in $\mathbb{R}^k$, with its bottom-left corner at $z_1^k$ and with side length $a$. For example, we have that $\qRVm{X_1^k}=z_1^k$ if and only if $X_1^k\in\dom{C}(z_1^k,1/m)$.

Let $H(\cdot)$, $h(\cdot)$, and $\kld{\cdot}{\cdot}$ denote entropy, differential entropy, and relative entropy, respectively, and let $\mutinf{\cdot;\cdot}$ denote the mutual information~\cite{Cover_Information1}. We take logarithms to base $e\approx 2.718$, so mutual informations and entropies have dimension \emph{nats}. The entropy rate of a discrete-valued, stationary $L$-variate stochastic process $\{\Xb_t\}$ is~\cite[Th.~4.2.1]{Cover_Information1}
\begin{equation}
\label{eq:entrate}
\entrate{\{\Xb_t\}} \triangleq \lim_{k\to\infty} \frac{H(\Xb_1^k)}{k}.
\end{equation}

\subsection{Information Dimension of RVs}

R\'enyi defined the information dimension of a collection of RVs $X_\ell^k$ as~\cite{Renyi_InfoDim}
\begin{equation}\label{eq:infodim}
 \infodim{X_\ell^k} \triangleq \limm \frac{\ent{\qRVm{X_\ell^k}}}{\log m}
\end{equation}
provided the limit exists. If the limit does not exist, one can define the upper and lower information dimension $\overline{d}(X_\ell^k)$ and $\underline{d}(X_\ell^k)$ by replacing the limit with the limit superior and limit inferior, respectively. If a result holds for both the limit superior and the limit inferior but it is unclear whether the limit exists, then we shall write $\overline{\underline{d}}(X_\ell^k)$. We shall follow this notation throughout this document: an overline $\overline{(\cdot)}$ indicates that the quantity in the brackets has been computed using the limit superior over $m$, an underline $\underline{(\cdot)}$ indicates that it has been computed using the limit inferior, both an overline and an underline $\overline{\underline{(\cdot)}}$ indicates that a result holds irrespective of whether the limit superior or limit inferior over $m$ is taken.

If $\ent{\qRV{X_\ell^k}{1}}<\infty$, then~\cite[Eq.~7]{Renyi_InfoDim},~\cite[Prop.~1]{Wu_Renyi} 
 \begin{equation}\label{eq:dim:bound}
 0\le \underline{d}(X_\ell^k) \le \overline{d}(X_\ell^k)\le k-\ell+1.
\end{equation}
If $\ent{\qRV{X_\ell^k}{1}}=\infty$, then $\overline{\underline{d}}(X_\ell^k)=\infty$. As shown in \cite[Lemma~3]{Wu_PhD}, information dimension is invariant under scaling and translation, i.e., $\overline{\underline{d}}(a\cdot X_\ell^k)=\overline{\underline{d}}(X_\ell^k)$ and $\overline{\underline{d}}(X_\ell^k+c)=\overline{\underline{d}}(X_\ell^k)$ for every  $a\neq 0$ and $c\in\mathbb{R}^{k-\ell+1}$.

\section{Information Dimension\\ of Univariate Processes}\label{sec:scalar}
In~\cite{GeigerKoch_ISIT,GeigerKoch_DimRate}, we generalized~\eqref{eq:infodim} by defining the information dimension rate of a univariate stationary process $\{X_t\}$ as
\begin{equation}\label{eq:dimrate}
 d(\{X_t\}) \triangleq \limm \frac{\entrate{\{\qRVm{X_t}\}}}{\log m} = \limm\limk \frac{\ent{\qRVm{X_1^k}}}{k\log m}
\end{equation}
provided the limit exists. (The limit over $k$ exists by stationarity.)

If $\ent{\qRV{X_1}{1}}<\infty$, then~\cite[Lemma~4]{GeigerKoch_DimRate}
 \begin{equation}\label{eq:dimrate:bound}
 0\le \underline{d}(\{X_t\})  \le \overline{d}(\{X_t\}) \le 1;
\end{equation}
if $\ent{\qRV{X_1}{1}}=\infty$, then $\overline{\underline{d}}(\{X_t\}) =\infty$. Moreover, the information dimension rate of the process cannot exceed the information dimension of the marginal RV, i.e.,
\begin{equation}
  \overline{\underline{d}}(\{X_t\})\le \overline{\underline{d}}(X_1).
\end{equation}

Kawabata and Dembo \cite[Lemma~3.2]{Kawabata_RDDim} showed that the information dimension of a RV equals its rate-distortion dimension. By emulating the proof of~\cite[Lemma~3.2]{Kawabata_RDDim}, we generalized this result to stationary processes by demonstrating that the information dimension rate is equal to the rate-distortion dimension. Specifically, let $R(X_1^k,D)$ denote the rate-distortion function of the $k$-dimensional source $X_1^k$, i.e.,
\begin{equation}\label{eq:R2(D)}
R(X_1^k,D) \triangleq \inf_{\mathsf{E}[\|\hat{X}_1^k-X_1^k\|^2]\leq D} \mutinf{X_1^k;\hat{X}_1^k}
\end{equation}
where the infimum is over all conditional distributions of $\hat{X}_1^k$ given $X_1^k$ such that $\mathsf{E}[\|\hat{X}_1^k-X_1^k\|^2]\leq D$ (where $\|\cdot\|$ denotes the Euclidean norm). The rate-distortion dimension of the stationary process $\{X_t\}$ is defined as
\begin{equation}\label{eq:rddim:def}
 \mathsf{dim}_R(\{X_t\}) \triangleq  2\lim_{D\downarrow 0} \lim_{k\to\infty} \frac{R(X_1^k,kD)}{-k\log D}
\end{equation}
provided the limit as $D\downarrow 0$ exists. By stationarity, the limit over $k$ always exists \cite[Th.~9.8.1]{gallager68}. We showed that \cite[Th.~5]{GeigerKoch_DimRate}
\begin{equation}\label{eq:rddim:stationary}
 \overline{\underline{\mathsf{dim}}}_R(\{X_t\})=\overline{\underline{d}}(\{X_t\}).
\end{equation}
This result directly generalizes to non-stationary process (possibly with the limit over $k$ replaced by the limit superior or limit inferior).

\section{Information Dimension\\ of Multivariate Processes}\label{sec:multivariate}
In this section, we generalize the definition of the information dimension rate \eqref{eq:dimrate} to multivariate (real-valued) processes and study its properties.

\begin{definition}[Information Dimension Rate]\label{def:infodim}
The information dimension rate of the $L$-variate, stationary process $\{\Xb_t\}$ is
\begin{align}
 \dimpinrate 
&\triangleq \limm \frac{\entrate{\{[\dom{\Xb}_t]_m\} }}{\log m} \notag\\
&= \limm \limk \frac{\ent{\qRVm{X_{1,1}^k},\dots ,\qRVm{X_{L,1}^k}}}{k \log m}
\end{align}
provided the limit over $m$ exists. 
\end{definition}

We next summarize some basic properties of the information dimension rate.

\begin{lem}[Finiteness and Bounds]\label{lem:bounds}
 Let $\{\Xb_t\}$ be a stationary, $L$-variate process. If $\ent{\qRV{\Xb_1}{1}}<\infty$, then
 \begin{equation}
 \label{eq:bounds}
   0 \le \overline{\underline{d}}(\{\Xb_t\}) \le \overline{\underline{d}}(\Xb_1) \le L.
    \end{equation}
If $\ent{\qRV{\Xb_1}{1}}=\infty$, then $\overline{\underline{d}}(\{\Xb_t\}) =\infty$.
\end{lem}

\begin{IEEEproof}
 Suppose first that $\ent{\qRV{\Xb_1}{1}}<\infty$. Then, the rightmost inequality in \eqref{eq:bounds} follows from~\eqref{eq:dim:bound}. The leftmost inequality follows from the nonnegativity of entropy. Finally, the center inequality follows since conditioning reduces entropy, hence $\entrate{\{\qRVm{\Xb_t}\}} \le \ent{\qRVm{\Xb_1}}$.
 
 Now suppose that $\ent{\qRV{\Xb_1}{1}}=\infty$. By stationarity and since $\qRV{\Xb_1}{1}$ is a function of $\qRVm{\Xb_1^k}$ for every $m$ and every $k$, we have 
\begin{equation}
 \ent{\qRV{\Xb_1}{1}} \le \ent{\qRV{\Xb_1^k}{m}}.
\end{equation}
This implies that $\entrate{\{\qRVm{\Xb_t}\}}=\infty$ and the claim $\overline{\underline{d}}(\{\Xb_t\}) =\infty$ follows from Definition~\ref{def:infodim}.
\end{IEEEproof}

It was shown in \cite[Lemma~3]{Wu_PhD} that information dimension is invariant under scaling and translation. The same properties hold for the information dimension rate. 

\begin{lem}[Scale Invariance]\label{lem:scaleinv}
 Let $\{\Xb_t\}$ be a stationary, $L$-variate process and let $a_i>0$, $i=1,\dots,L$. Further let $Y_{i,t}\triangleq a_i X_{i,t}$, $i=1,\dots,L$, $t\in\mathbb{Z}$. Then, $\overline{\underline{d}}(\{\Yb_t\})=\overline{\underline{d}}(\{\Xb_t\})$.
\end{lem}

\begin{IEEEproof}
We show the proof for $L=2$ by adapting the proof of~\cite[Lemma~16]{Wu_Renyi}. The proof for higher-dimensional processes follows along the same lines. By the data processing inequality, the chain rule, and the fact that conditioning reduced entropy, we have
 \begin{align}
  &\ent{\qRVm{a_1X_{1,1}^k},\qRVm{a_2X_{2,1}^k}}\notag \\
 &\le \ent{\qRVm{a_1X_{1,1}^k},\qRVm{a_2X_{2,1}^k},\qRVm{X_{1,1}^k},\qRVm{X_{2,1}^k}}\notag\\
 & = \ent{\qRVm{X_{1,1}^k},\qRVm{X_{2,1}^k}} \notag \\
 & \quad + \ent{\qRVm{a_1X_{1,1}^k},\qRVm{a_2X_{2,1}^k}\bigm|\qRVm{X_{1,1}^k},\qRVm{X_{2,1}^k}}\notag\\
 &\le \ent{\qRVm{X_{1,1}^k},\qRVm{X_{2,1}^k}}+ \ent{\qRVm{a_1X_{1,1}^k}\bigm|\qRVm{X_{1,1}^k}}\notag\\
 & \quad + \ent{\qRVm{a_2X_{2,1}^k}\bigm|\qRVm{X_{2,1}^k}}\label{eq:proofscaleinv:bound1}
 \end{align}
 Now let $z_1^k\in\mathbb{Z}^k$. We have
 \begin{align}
  &\ent{[a_1X_{1,1}^k]_m\Biggm|[X_{1,1}^k]_m=\frac{z_1^k}{m}} \notag\\
  &= \ent{[a_1X_{1,1}^k]_m\Bigm|X_{1,1}^k\in \textstyle\dom{C}(\frac{z_1^k}{m},\frac{1}{m})}\notag\\
  &= \ent{\lfloor a_1mX_{1,1}^k\rfloor\Bigm|a_1mX_{1,1}^k\in\textstyle\dom{C}(a_1 z_1^k,a_1)}\notag\\
  &\le k \log \left(\lceil a_1 \rceil+1\right).\label{eq:proofscaleinv:bound2}
 \end{align}
 where the last inequality follows because flooring numbers from an interval of length $a_1$ yields at most $\lceil a_1 \rceil+1$ different integers. Combining~\eqref{eq:proofscaleinv:bound2} with~\eqref{eq:proofscaleinv:bound1}, dividing by $k$, and letting $k$ tend to infinity yields
 \begin{multline}
  \entrate{\{\qRVm{a_1X_{1,t}}\},\{\qRVm{a_2X_{2,t}}\}}  \\
 \le \entrate{\{\qRVm{X_{1,t}}\},\{\qRVm{X_{2,t}}\}}\\
 \quad + \log(\lceil a_1\rceil +1)+ \log(\lceil a_2\rceil +1).
 \end{multline}
 Dividing by $\log m$ and letting $m$ tend to infinity yields $d(\{\Yb_t\})\le\dimpinrate$. The reverse inequality is obtained by noting that $\frac{1}{a_1}a_1X_{1,t} = X_{1,t}$, so the above steps with $(a_1,a_2)$ replaced by $(1/a_1,1/a_2)$ yield
 \begin{multline*}
  \entrate{\{\qRVm{a_1X_{1,t}}\},\{\qRVm{a_2X_{2,t}}\}}  \\
 \ge \entrate{\{\qRVm{X_{1,t}}\},\{\qRVm{X_{2,t}}\}}\\
 \quad - \log(\lceil 1/a_1\rceil +1)- \log(\lceil 1/a_2\rceil +1).
 \end{multline*}
Thus, we obtain $d(\{\Yb_t\})\ge\dimpinrate$ by dividing by $\log m$ and by letting $m$ tend to infinity.
\end{IEEEproof}

\begin{lem}[Translation Invariance]\label{lem:rateTransInv}
  Let $\{\Xb_t\}$ be a stationary, $L$-variate process and let $\{\mathbf{c}_t\}$, $t\in\mathbb{Z}$ be a sequence of $L$-dimensional vectors. Then, $\overline{\underline{d}}(\{\Xb_t+\mathbf{c}_t\}) = \overline{\underline{d}}(\{\Xb_t\})$.
\end{lem}

\begin{IEEEproof}
The lemma follows from \cite[Lemma~30]{Wu_PhD}, which states that
\begin{equation}\label{eq:transbound}
 |\ent{U_1^{kL}} -\ent{V_1^{kL}} | \le \sum_{i=1}^{kL} \log(1+A_i+B_i)
\end{equation}
for any collection of integer-valued RVs $U_1^{kL}$ and $V_1^{kL}$ satisfying almost surely $-B_i\le U_i-V_i\le A_i$, $i=1,\dots,kL$. Applying this result with $U_{\ell L+j}=\lfloor mX_{\ell,j} + m c_{\ell,j} \rfloor$ and $V_{\ell L+j}=\lfloor mX_{\ell,j} \rfloor + \lfloor m c_{\ell,j} \rfloor$ gives the desired result. Indeed, we have that $-1 \le U_{\ell L+j}-V_{\ell L+j}\le 2$, so \eqref{eq:transbound} yields
\begin{equation}
\label{eq:transbound_2}
 \left|\ent{[\Xb_1^k]_m}-\ent{[\Xb_1^k+\mathbf{c}_1^k]_m} \right| \le kL\log(4).
\end{equation}
We thus obtain $|\dimpinrate-d(\{\Xb_t+\mathbf{c}_t\})| = 0$ by dividing \eqref{eq:transbound_2} by $k\log m$ and by letting $k$ and $m$ tend to infinity.
\end{IEEEproof}

We finally observe that the information dimension rate of a stationary stochastic process equals its rate-distortion dimension. This generalizes~\cite[Th.~5]{GeigerKoch_DimRate} to multivariate processes.

\begin{thm}\label{lem:RDDim}
 Let $\{\Xb_t\}$ be an $L$-variate stationary process. Then,
 \begin{equation}\label{eq:rddim:mult}
  \overline{\underline{d}}(\{\Xb_t\}) = \overline{\underline{\mathsf{dim}}}_R{\{\Xb_t\}}
 \end{equation}
 where $\mathsf{dim}_R\{\Xb_t\}$ is defined as in \eqref{eq:rddim:def} but with $\{X_t\}$ replaced by $\{\Xb_t\}$.
\end{thm}
\begin{IEEEproof}
The proof is analog to that of \cite[Lemma~3.2]{Kawabata_RDDim} and \cite[Th.~5]{GeigerKoch_DimRate}. We thus only provide a short sketch. We have with $m=1/\sqrt{D}$ and from~\cite[Eq.~(71)~\&~(81)]{GeigerKoch_DimRate} that
 \begin{multline}
  \ent{\qRVm{\Xb_1^k}} - 2k\left(1+\sum_{i=0}^\infty \e{-i^2}\right)\\ \le R(\Xb_1^k,kLD) \le \ent{\qRVm{\Xb_1^k}}.
 \end{multline}
 Dividing by $k$ and taking the limit inferior and limit superior as $k\to\infty$, we get
 \begin{IEEEeqnarray}{lCl}
  \IEEEeqnarraymulticol{3}{l}{\entrate{\{\qRVm{\Xb_t}\}} -2\left(1+\sum_{i=0}^\infty \e{-i^2}\right)} \nonumber\\
\qquad & \le & \lim_{k\to \infty} \frac{R(\Xb_1^k,kLD)}{k} \nonumber\\
& \le & \entrate{\{\qRVm{\Xb_t}\}}
 \end{IEEEeqnarray}
 where the limit exists because $\{\Xb_t\}$ is stationary. We now divide the outer terms of this inequality by $\log m$ and the inner term by $-\frac{1}{2}\log D$, and take the limits as $m\to\infty$ and $D\downarrow 0$, respectively. This yields
\begin{equation}
 \overline{\underline{d}}(\{\Xb_t\}) \le \overline{\underline{\mathsf{dim}}}_R{\{\Xb_t\}} \le\overline{\underline{d}}(\{\Xb_t\})
\end{equation}
and proves~\eqref{eq:rddim:mult}.
 \end{IEEEproof}

\section{Information Dimension\\ of Gaussian Processes}\label{sec:GaussMult}
Let $\{\Xb_t\}$ be a stationary, $L$-variate, real-valued Gaussian process with mean vector $\boldsymbol{\mu}$ and (matrix-valued) spectral distribution function (SDF) $\theta\mapsto \sdft{\Xb}$. Thus, $\sdf{\Xb}$ is bounded, non-decreasing, and right-continuous on $[-1/2,1/2]$, and it satisfies \cite[(7.3), p.~141]{Wiener_StochasticProcesses}
\begin{equation}
\covt{\Xb} = \int_{-1/2}^{1/2} \e{-\imath 2\pi\tau\theta} \diff \sdft{\Xb}, \quad \tau\in\mathbb{Z}
\end{equation}
where $\covt{\Xb} \triangleq \expec{(\Xb_{t+\tau}-\boldsymbol{\mu})\trans{(\Xb_t-\boldsymbol{\mu})}}$ denotes the autocovariance function. It can be shown that $\theta\mapsto \sdft{\Xb}$ has a derivative almost everywhere, which has positive semi-definite, Hermitian values \cite[(7.4), p.~141]{Wiener_StochasticProcesses}. We shall denote the derivative of $\sdf{\Xb}$ by $\dsdf{\Xb}$. Further note that the $(i,j)$-th element of $\sdf{\Xb}$ is the cross SDF $\theta\mapsto\sdf{X_iX_j}$ of the component processes $\{X_{i,t}\}$ and $\{X_{j,t}\}$, i.e.,
\begin{equation}
\covt{X_iX_j} = \int_{-1/2}^{1/2} \e{-\imath 2\pi\tau\theta} \diff \sdft{X_iX_j}, \quad \tau\in\mathbb{Z}
\end{equation}
where $\covt{X_iX_j} \triangleq \expec{(X_{i,t+\tau}-\mu_i)(X_{j,t}-\mu_j)}$ denotes the cross-covariance function.

For univariate stationary Gaussian processes with SDF $\sdf{X}$, we have shown that the information dimension rate is equal to the Lebesgue measure of the set of harmonics on $[-1/2,1/2]$ where the derivative of $\sdf{X}$ is is positive \cite[Th.~7]{GeigerKoch_DimRate}, i.e.,
\begin{equation}\label{eq:LebesgueSupport}
 d(\{X_t\}) = \lambda(\{\theta{:}\ \dsdft{X}>0\}).
\end{equation}

This result can be directly generalized to the multivariate case where the component processes are independent. Indeed, suppose that $\{\Xb_t\}$ is a collection of $L$ independent Gaussian processes $\{X_{i,t},\,t\in\mathbb{Z}\}$ with SDFs $\sdf{X_i}$. This corresponds to the case where the (matrix-valued) SDF is a diagonal matrix with the SDFs of the individual processes on the main diagonal. For independent processes, the joint entropy rate can be written as the sum of the entropy rates of the component processes. It follows that
\begin{equation}
\label{eq:idrG_indep}
\dimpinrate= \sum_{i=1}^L d(\{X_{i,t}\}) = \sum_{i=1}^L\lebesgue(\{\theta \colon \dsdft{X_i}>0\}).
\end{equation}
The expression on the right-hand side (RHS) of \eqref{eq:idrG_indep} can alternatively be written as
\begin{equation}
\label{eq:idrG_indep_2}
\int_{-1/2}^{1/2}  \sum_{i=1}^L\mathbf{1}\{\dsdft{X_i}>0\} \diff \theta = \int_{-1/2}^{1/2}  \mathrm{rank}(\dsdft{\Xb}) \diff \theta
\end{equation}
where $\mathbf{1}\{\cdot\}$ is the indicator function. Observe that it is immaterial at which frequencies the component processes contain signal power. For example, the information dimension rate of two independent Gaussian processes with bandwidth $1/4$ equals $1$ regardless of where the derivatives of their SDFs have their support.
The following theorem shows that this result continuous to hold for general $L$-variate Gaussian processes.

\begin{thm}\label{thm:multivariate}
 Let $\{\Xb_t\}$ be a stationary, $L$-variate Gaussian process with mean vector $\bfmu$ and SDF $\sdf{\Xb}$. Then,
 \begin{equation}\label{eq:thm:mainline}
  \dimpinrate = \int_{-1/2}^{1/2}  \mathrm{rank}(\dsdft{\Xb}) \diff \theta.
 \end{equation}
\end{thm}

\begin{IEEEproof}
See Appendix~\ref{app:main}.
\end{IEEEproof}

An important ingredient in the proof is the following generalization of~\cite[Lemma~6]{GeigerKoch_DimRate}.

\begin{lem}\label{lem:quantizedProcess}
 Let $\{\Xb_t\}$ be an $L$-variate, stationary Gaussian process with mean vector $\bfmu$ and SDF $\sdft{\Xb}$. Then, the $(i,j)$-th entry of the SDF $\theta\mapsto\sdft{\qRVm{\Xb}}$ of $\{\qRVm{\Xb_t}\}$, i.e., the  satisfies for $i,j=1,\dots,L$,
  \begin{equation}\label{eq:SDMSum}
  \sdft{\qRVm{X_i}\qRVm{X_j}} = (a_i + a_j - 1) \sdft{X_iX_j} + \sdft{N_iN_j}
 \end{equation}
 where $N_{i,t}\triangleq X_{i,t}-\qRVm{X_{i,t}}$ 
 and where
\begin{equation}
 a_i\triangleq\frac{1}{\sigma_i^2}\expec{(X_{j,t}-\mu_j)(\qRVm{X_{i,t}}-\expec{\qRVm{X_{i,t}}})}.
\end{equation}
For every $i=1,\dots,L$, we have
 \begin{equation}\label{eq:boundBussgang}
  |1-a_i| \le \frac{1}{m}\sqrt{\frac{2}{\pi\sigma_i^2}}
 \end{equation}
 and
 \begin{equation}\label{eq:boundNoise}
  \int_{-1/2}^{1/2} \diff\sdft{N_i} \le \frac{1}{m^2}.
 \end{equation}
Moreover, if all component processes have unit variance, then $a_1=\ldots=a_L$ and hence
\begin{equation}\label{eq:SDMSumEqual}
 \sdft{\qRVm{\Xb}} = (2a_1-1)\sdft{\Xb} + \sdft{\Nb}.
\end{equation}
\end{lem}

\begin{IEEEproof}
Let $Z_{i,t}\triangleq \qRV{X_{i,t}}{m}$. For every pair $i,j=1,\dots,L$, we have with~\cite[(83)]{GeigerKoch_DimRate}
\begin{align}
 \covt{N_iN_j} = & \covt{X_iX_j} + \covt{Z_iZ_j}\nonumber\\ &{} - \covt{X_iZ_j} - \covt{Z_iX_j}.
\end{align}
From Bussgang's theorem~\cite[eq.~(19)]{bussgang52} we have that $\covt{X_iZ_j} = \cov{Z_jX_i}(-\tau)= a_{j}  \covt{X_iX_j}$. This implies that
 \begin{align}
  \covt{N_iN_j} 
  &=\covt{X_iX_j} +  \covt{Z_iZ_j}\notag\\
 &\quad - a_j\covt{X_iX_j} - a_i\cov{X_jX_i}(-\tau)\notag\\
  &=  (1-a_j-a_i)\covt{X_iX_j} + \covt{Z_iZ_j}.
 \end{align}
 Since the SDM is fully determined by the covariance structure of a process, we obtain~\eqref{eq:SDMSum}. Equations~\eqref{eq:boundBussgang} and~\eqref{eq:boundNoise} follow immediately from equations (35) and (36) in~\cite[Lemma~6]{GeigerKoch_DimRate}.
\end{IEEEproof}

\section{Information Dimension\\ of Complex Gaussian Processes}\label{sec:complex}

Theorem~\ref{thm:multivariate} allows us to study the information dimension of stationary, univariate, complex-valued Gaussian processes by treating them as bivariate, real-valued processes. Let $\{Z_t\}$ be a stationary, univariate, complex-valued, Gaussian process with mean $\mu$ and SDF $\sdf{Z}$, i.e., 
\begin{equation}
\covt{Z}=\int_{-1/2}^{1/2} e^{-\imath 2 \pi \tau \theta}\diff\sdft{Z}, \quad \tau\in\mathbb{Z}
\end{equation}
where $\covt{Z}\triangleq\expec{(Z_{t+\tau}-\mu)(Z_t-\mu)^*}$ is the autocovariance function.

Alternatively, $\{Z_t\}$ can be expressed in terms of its real and imaginary part. Indeed, let $Z_t=R_t+\imath I_t$, $t\in\mathbb{Z}$. The stationary, bivariate, real-valued process $\{(R_t,I_t),\,t\in\mathbb{Z}\}$ is jointly Gaussian and has SDF
\begin{equation}
\sdft{(R,I)} = \left(\begin{array}{cc} \sdft{R} & \sdft{RI} \\ \sdft{IR} & \sdft{I} \end{array}\right), \quad -1/2\leq\theta\leq 1/2
\end{equation}
where $\sdf{R}$ and $\sdf{I}$ are the SDFs of $\{R_t\}$ and $\{I_t\}$, respectively, and $\sdf{RI}$ and $\sdf{IR}$ are the cross SDFs between $\{R_t\}$ and $\{I_t\}$. The derivatives of $\sdf{Z}$ and $\sdf{(R,I)}$ are connected as follows:
\begin{IEEEeqnarray}{lCl}
 \dsdft{Z} & = & \dsdft{R}+ \dsdft{I}+\imath\bigl(\dsdft{IR}-\dsdft{RI}\bigr) \nonumber\\
& = & \dsdft{R}+ \dsdft{I}+2\mathfrak{Im}\bigl(\dsdft{RI}\bigr)  \label{eq:prooflem:FZ}
\end{IEEEeqnarray}
where the last equality follows because $\dsdf{(R,I)}$ is Hermitian. It can be further shown that $\theta\mapsto \dsdft{R}$ and $\theta\mapsto\dsdft{I}$ are real-valued and symmetric, and that $\theta\mapsto \mathfrak{Im}\bigl(\dsdft{RI}\bigr)$ is anti-symmetric.

A stationary, complex-valued process $\{Z_t\}$ is said to be \emph{proper} if its mean $\mu$ and its pseudo-autocovariance function
\begin{equation*}
 \pcovt{Z} \triangleq\expec{(Z_{t+\tau}-\mu)(Z_t-\mu)}, \quad \tau\in\mathbb{Z}
\end{equation*}
are both zero~\cite[Def.~17.5.4]{Lapidoth_DigitalCommunication}. Since, by Lemma~\ref{lem:rateTransInv}, the information dimension rate is independent of $\mu$, we shall slightly abuse notation and say that a stationary, complex-valued process is proper if its pseudo-autocovariance function is identically zero, irrespective of its mean. Properness implies that, for all $\theta$, $\sdft{R}=\sdft{I}$ and $\sdft{RI}=-\sdft{IR}$. Since $\theta\mapsto\dsdft{(R,I)}$ is Hermitian, this implies that for a proper process the derivative of the cross SDF $\sdf{RI}$ is purely imaginary.

The following corollary to Theorem~\ref{thm:multivariate} shows that proper Gaussian processes maximize information dimension. This parallels  the result that proper Gaussian vectors maximize differential entropy~\cite[Th.~2]{Neeser_Proper}.
\begin{cor}\label{cor:complex}
 Let $\{Z_t\}$ be a stationary, complex-valued Gaussian process with mean $\mu$ and SDF $\sdf{Z}$. Then
\begin{equation}\label{eq:upperbound}
 d(\{Z_t\}) \le 2\cdot\lambda(\{\theta{:}\ \dsdft{Z}>0\})
\end{equation}
with equality if $\{Z_t\}$ is proper.
\end{cor}
\begin{IEEEproof}
 We know from Theorem~\ref{thm:multivariate} that
 \begin{equation}
 \label{eq:prooflem:complex:1}
 d(\{Z_t\})= \int_{-1/2}^{1/2}  \mathrm{rank}(\dsdft{(R,I)}) \diff \theta.
 \end{equation}
 For a given $\theta$, the eigenvalues of $\dsdft{(R,I)}$ are given by
\begin{equation}
 \frac{\dsdft{R}+\dsdft{I}}{2} \pm\sqrt{\frac{(\dsdft{R}-\dsdft{I})^2}{4} + |\dsdft{RI}|^2}.
\end{equation}
Since $\dsdft{(R,I)}$ is positive semi-definite, these eigenvalues are nonnegative and 
\begin{equation}\label{eq:posdef}
 \dsdft{R}\dsdft{I}\ge|\dsdft{RI}|^2.
\end{equation}
In particular, the larger of these eigenvalues, say $\mu_1(\theta)$, is zero on
\begin{equation}
\dom{F}_1\triangleq\{\theta\colon \dsdft{R}=\dsdft{I}=0\}.
\end{equation}
The smaller eigenvalue, $\mu_2(\theta)$, is zero on
\begin{equation}
\dom{F}_2\triangleq \bigl\{\theta\colon\dsdft{R}\dsdft{I}=|\dsdft{RI}|^2\bigr\}.
\end{equation}
Clearly, $\dom{F}_1\subseteq \dom{F}_2$. By \eqref{eq:prooflem:complex:1}, we have that
\begin{align}
 d(\{Z_t\}) 
&= \lambda(\{\theta{:}\ \mu_1(\theta)>0\}) + \lambda(\{\theta{:}\ \mu_2(\theta)>0\})\notag\\
&= 1-\lambda(\dom{F}_1) +1-\lambda(\dom{F}_1)  -\lambda(\dom{F}_1^\mathsf{c}\cap\dom{F}_2). \label{eq:prooflem:d(Z)_1}
\end{align}

We next note that, by \eqref{eq:prooflem:FZ} and~\eqref{eq:posdef}, the derivative $\dsdft{Z}$ is zero if either $\dsdft{R}=\dsdft{I}=0$ or if $\dsdft{R}+\dsdft{I}>0$ and $\dsdft{R}+\dsdft{I}=-2\mathfrak{Im}(\dsdft{RI})$. Since $\theta\mapsto\dsdft{R}$ and $\theta\mapsto\dsdft{I}$ are symmetric and $\theta\mapsto\mathfrak{Im}(\dsdf{RI})$ is anti-symmetric, it follows that for any $\theta\in\dom{F}_1^c$ satisfying $\dsdft{R}+\dsdft{I}=-2\mathfrak{Im}(\dsdft{RI})$ we have that $\dsdf{R}(-\theta)+\dsdf{I}(-\theta)=2\mathfrak{Im}(\dsdf{RI}(-\theta))$. Thus, defining
\begin{equation}
\dom{F}_3 \triangleq \bigl\{\theta\colon \dsdft{R}+ \dsdft{I}=2|\mathfrak{Im}(\dsdft{RI})|\bigr\}
\end{equation}
we can express the Lebesgue measure of the set of harmonics where $\dsdft{Z}=0$ as
\begin{equation}
\label{eq:prooflem:d(Z)_2}
 \lambda(\{\theta{:}\ \dsdft{Z}=0\}) 
=\lambda(\dom{F}_1)+\frac{1}{2}\lambda(\dom{F}_1^{\mathsf{c}}\cap \dom{F}_3).
\end{equation}
Combining \eqref{eq:prooflem:d(Z)_1} and \eqref{eq:prooflem:d(Z)_2}, we obtain
\begin{equation}
d(\{Z_t\}) = 2 \lambda(\{\theta\colon \dsdft{Z}>0\}) + \lambda(\dom{F}_1^{\mathsf{c}}\cap\dom{F}_3) - \lambda(\dom{F}_1^{\mathsf{c}}\cap\dom{F}_2). \label{eq:prooflem:bla}
\end{equation}
Since the arithmetic mean is greater than or equal to the geometric mean, and using \eqref{eq:posdef}, we have that
\begin{multline}
 (\dsdft{R}+ \dsdft{I})^2 \ge 4\dsdft{R}\dsdft{I}\\
 \ge 4|\dsdft{RI}|^2 \ge 4\mathfrak{Im}(\dsdft{RI})^2.
\end{multline}
Hence, $\dom{F}_3\subseteq \dom{F}_2$ which implies that the difference of the last two Lebesgue measures on the RHS of \eqref{eq:prooflem:bla} is less than or equal to zero. This proves \eqref{eq:upperbound}.

If $\{Z_t\}$ is proper, then we have $\dsdft{R}=\dsdft{I}$ and $|\dsdft{RI}|=|\mathfrak{Im}(\dsdft{RI})|$ for almost all $\theta$. In this case, $\dsdft{R}\dsdft{I}=|\dsdft{RI}|^2$ implies $\dsdft{R}+ \dsdft{I}=2|\mathfrak{Im}(\dsdft{RI})|$, so $\dom{F}_2\subseteq\dom{F}_3$. It follows that $\dom{F}_2=\dom{F}_3$ and the difference of the last two Lebesgue measures on the RHS of \eqref{eq:prooflem:bla} is zero. Hence, \eqref{eq:upperbound} holds with equality.
\end{IEEEproof}

\begin{remark}
There are also non-proper processes for which \eqref{eq:upperbound} holds with equality. For example, this is the case for any stationary Gaussian process for which real and imaginary parts are independent, i.e., $\dsdft{RI}=0$, and $\dsdf{R}$ and $\dsdf{I}$ have matching support but are different otherwise.
\end{remark}

\section{Conclusion}
We proposed a generalization of information dimension to multivariate, stationary processes. Specifically, if $\{\Xb_t\}$ is an $L$-variate, stationary process, then we defined its information dimension rate $d(\{\Xb_t\})$ as the entropy rate $\entrate{\qRVm{\Xb_t}}$ divided by $\log m$ in the limit as $m\to\infty$. We demonstrated that the information dimension rate is bounded if $\qRV{\Xb_1}{1}$ has finite entropy and that it is invariant under scaling and translation. We furthermore showed that $d(\{\Xb_t\})$ coincides with the rate-distortion dimension, thus generalizing our result for univariate processes~\cite[Th.~5]{GeigerKoch_DimRate}.

Our main result concerns the information dimension rate of $L$-variate, stationary, Gaussian processes with (matrix-valued) spectral distribution function $\sdf{\Xb}$. We showed that in this case $d(\{\Xb_t\})$ equals the Lebesgue integral of the rank of the derivative of $\sdf{\Xb}$, i.e.,
\begin{equation}
 \dimpinrate = \int_{-1/2}^{1/2}  \mathrm{rank}(\dsdft{\Xb}) \diff \theta.
\end{equation}
As a corollary, we showed that the information dimension rate of a univariate complex-valued Gaussian process $\{X_t\}$ is upper-bounded by twice the Lebesgue measure of the support of the derivative of its spectral distribution function $\sdf{X}$. This upper bound is achieved if, but not only if, the Gaussian process is proper.

\appendices
\section{Proof of the Main Result}\label{app:main}
The proof follows along the lines of~\cite[Th.~7]{GeigerKoch_DimRate}. We adopt the following notation. For every $i$, we define $N_{i,t}\triangleq X_{i,t}-\qRVm{X_{i,t}}$. Let $U_{i,t}$ be i.i.d. (over all $i$ and $t$) and uniformly distributed on $[0,1/m)$, and let $W_{i,t} \triangleq \qRVm{X_{i,t}}+U_{i,t}$. We define $\{\qRVm{\Xb_{t}}\}$, $\{\Nb_t\}$, and $\{\Ub_t\}$ as corresponding multivariate processes. Since $\{U_{i,t}\}$ is independent of $\{\qRVm{X_{j,t}}\}$ for every $i,j$, the (matrix-valued) SDFs of $\{\Wb_t\}$, $\{\qRVm{\Xb_{t}}\}$, and $\{\Ub_t\}$ satisfy
\begin{equation}
\sdft{\Wb}=\sdft{\qRVm{\Xb}}+\sdft{\Ub}, \quad -1/2 \leq \theta \leq 1/2.
\end{equation}
Moreover, the (matrix-valued) power spectral density (PSD) of $\{\Ub_t\}$ exists and equals $\psdt{\Ub}=\frac{1}{12m^2} I_L$, where $I_L$ denotes the $L\times L$ identity matrix.

We next note that since the information dimension rate is translation invariant (Lemma~\ref{lem:rateTransInv}) and since the SDF $\sdf{\Xb}$ does not depend on the mean vector $\bfmu$, we can assume without loss of generality that $\{\Xb_t\}$ has zero mean. We moreover show in Lemma~\ref{lem:zerovariance} in Appendix~\ref{app:aux} that we can assume, without loss of generality, that every component process of $\{\Xb_t\}$ has unit variance. Indeed, if all component processes have positive variance, then this follows immediately from Lemma~\ref{lem:scaleinv}. Lemma~\ref{lem:zerovariance} expands upon Lemma~\ref{lem:scaleinv} in that it shows that 1) normalizing component processes to unit variance does not affect the rank of $\sdf{\Xb}$ and 2) component processes with zero variance need not be considered in computing $d(\{\Xb_t\})$ or the rank of $\sdf{\Xb}$.

With this assumption and the notation introduced above, we write the entropy of $\qRVm{\Xb_1^k}$ in terms of a differential entropy, i.e.,
 \begin{equation}
  \ent{\qRVm{\Xb_1^k}} = \diffent{\Wb_1^k}+kL\log m.
 \end{equation}
Denoting by $(\Wb_1^k)_G$ a Gaussian vector with the same mean and covariance matrix as $\Wb_1^k$, and denoting by $\pdf{\Wb_1^k}$ and $\pdfg{\Wb_1^k}$ the probability density functions (pdfs) of $\Wb_1^k$ and $(\Wb_1^k)_G$, respectively, this can be expressed as
 \begin{equation}
 \label{eq:derate_quantent}
  H(\qRV{\Xb_1^k}{m}) = \diffent{(\Wb_1^k)_G} + \kld{\pdf{\Wb_1^k}}{\pdfg{\Wb_1^k}} + kL\log m.
 \end{equation}
Dividing by $k\log m$ and letting first $k$ and then $m$ tend to infinity yields the information dimension rate $d(\{\Xb_t\})$. Lemma~\ref{lem:KLDBounded} in Appendix~\ref{app:aux} shows that
\begin{equation}
\kld{\pdf{\Wb_1^k}}{\pdfg{\Wb_1^k}} \leq k\mathsf{K}
\end{equation}
for some constant $\mathsf{K}$ that is independent of $(k,m)$. Moreover, the differential entropy rate of the stationary, $L$-variate Gaussian process $\{\Wb_t\}$ is given by~\cite[Th.~7.10]{Wiener_StochasticProcesses}
 \begin{multline}\label{eq:derate_sdm}
  \limk \frac{\diffent{(\Wb_1^k)_G}}{k}  \\ = \frac{L}{2}\log (2\pi e) + \frac{1}{2} \int_{-1/2}^{1/2} \log\det\dsdft{\Wb} \diff\theta.
 \end{multline}
It thus follows that the information dimension rate of $\{\Xb_t\}$ equals
\begin{equation}\label{eq:proof:Lplus}
 d(\{\Xb_t\}) = L + \limm \frac{1}{2\log m} \int_{-1/2}^{1/2} \log\det\dsdft{\Wb} \diff\theta.
\end{equation}
It remains to show that the RHS of~\eqref{eq:proof:Lplus} is equal to the RHS of~\eqref{eq:thm:mainline}. To do so, we first show that the integral on the RHS of~\eqref{eq:proof:Lplus} can be restricted to a subset $\dom{F}_\Upsilon^\mathsf{c}\subseteq[-1/2,1/2]$ on which the entries of $\dsdft{\Nb}$ are bounded from above by $\Upsilon/m^2$ for some $\Upsilon>0$. We then show that, on this set, $\det\dsdft{\Wb}$ can be bounded from above and from below by products of affine transforms of the \emph{eigenvalues of $\dsdft{\Xb}$}. These bounds are asymptotically tight, i.e., they are equal in the limit as $m$ tends to infinity. We complete the proof by showing that the order of limit and integration can be exchanged.

\subsection{Restriction on $\dom{F}_\Upsilon^\mathsf{c}\subseteq[-1/2,1/2]$}
Choose $\Upsilon>0$ and let 
\begin{equation}
 \dom{F}_\Upsilon\triangleq \{\theta\colon \max_{i=1,\dots,L} \dsdft{N_{i}}> \Upsilon/m^2\}.
\end{equation}
We have from~\cite[(106)-(108)]{GeigerKoch_DimRate} that, for every $i$,
 \begin{equation}
  \lebesgue\left(\left\{\theta{:}\ \dsdft{N_{i}}> \frac{\Upsilon}{m^2}\right\}\right) \le \frac{1}{\Upsilon}.
 \end{equation}
  The set $\dom{F}_\Upsilon$ is the union of $L$ such events, from which $\lebesgue(\dom{F}_\Upsilon)\le\frac{L}{\Upsilon}$ follows from the union bound. Since
\begin{equation}
\label{eq:need_this}
\dsdft{\Wb} = \dsdft{\qRVm{\Xb}} + \psdt{\Ub}
\end{equation}
and since derivatives of matrix-valued SDFs are positive semidefinite, we have
\begin{equation}
\det \dsdft{\Wb} \ge \det \psdt{\Ub} = {1}/{(12m^2)^L}.
\end{equation}
Hence,
 \begin{IEEEeqnarray}{lCl}
  \IEEEeqnarraymulticol{3}{l}{\liminf_{m\to\infty} \frac{\int_{\dom{F}_\Upsilon} \log \det \dsdft{\Wb} \diff\theta}{\log m}} \nonumber\\
\qquad & \ge & - \lebesgue(\dom{F}_\Upsilon) \limm\frac{ 2L \log(12m)}{\log m} \nonumber\\
& \ge & -\frac{2L^2}{\Upsilon}
 \end{IEEEeqnarray}
 where $\liminf$ denotes the limit inferior. Here, the last step follows because $\lebesgue(\dom{F}_\Upsilon)\le\frac{L}{\Upsilon}$. Applying Hadamard's and Jensen's inequality we further get
 \begin{IEEEeqnarray}{lCl}
  \IEEEeqnarraymulticol{3}{l}{\int_{\dom{F}_\Upsilon} \log \det \dsdft{\Wb} \diff\theta} \nonumber\\
\qquad  &\le & \sum_{i=1}^L  \int_{\dom{F}_\Upsilon} \log \dsdft{W_{i}} \diff\theta\nonumber\\
  & \le & \sum_{i=1}^L\lebesgue(\dom{F}_\Upsilon) \log\left(\frac{\int_{\dom{F}_\Upsilon} \dsdft{W_{i}} \diff\theta}{\lebesgue(\dom{F}_\Upsilon)}\right)\nonumber\\
  &= &\sum_{i=1}^L\lebesgue(\dom{F}_\Upsilon) \log\left(\frac{(2a_1-1)+\frac{1}{12m^2}+\cov{N_{i}}(0)}{\lebesgue(\dom{F}_\Upsilon)}\right)\nonumber\\
  & = & L \lebesgue(\dom{F}_\Upsilon) \log\left(\frac{(2a_1-1)+\frac{1}{12m^2}+\frac{1}{m^2}}{\lebesgue(\dom{F}_\Upsilon)}\right)
 \end{IEEEeqnarray}
 where the last step is due to~\eqref{eq:boundNoise} in Lemma~\ref{lem:quantizedProcess}. Since, by~\eqref{eq:boundBussgang} in the same lemma, $a_1 \to 1$ with $m\to \infty$, we have
 \begin{equation}
  \limsup_{m\to\infty} \int_{\dom{F}_\Upsilon} \log \det \dsdft{\Wb} \diff\theta \le -L \lebesgue(\dom{F}_\Upsilon) \log\left(\lebesgue(\dom{F}_\Upsilon)\right)
\end{equation}
where $\limsup$ denotes the limit superior. As a consequence, we have 
  \begin{multline}
  -\frac{2L^2}{\Upsilon} 
\le \liminf_{m\to\infty} \frac{\int_{\dom{F}_\Upsilon} \log \det \dsdft{\Wb} \diff\theta}{\log m} \\
\le \limsup_{m\to\infty} \frac{\int_{\dom{F}_\Upsilon} \log \det \dsdft{\Wb} \diff\theta}{\log m} \label{eq:throw_away}
\le 0
 \end{multline}
 for every $\Upsilon$. It follows that this integral does not contribute to the information dimension rate if we let $\Upsilon$ tend to infinity. It thus suffices to evaluate the limit of
 \begin{equation}\label{eq:integralToEvaluate}
  L+\frac{1}{2\log m} \int_{\dom{F}^c_\Upsilon} \log \det \dsdft{\Wb} \diff\theta.
 \end{equation}

\subsection{Bounding $\det \dsdft{\Wb}$ by the Eigenvalues of $\dsdft{\Xb}$}
Lemma~\ref{lem:quantizedProcess} and \eqref{eq:need_this} yield
 \begin{equation}
  \dsdft{\Wb} = (2a_1-1)\dsdft{\Xb} + \dsdft{\Nb} + \frac{1}{12m^2} I_L.
 \end{equation}
Let $\mu_i(\theta)$, $i=1,\dots,L$, denote the eigenvalues of $\dsdft{\Xb}$. Sine $\dsdft{\Nb}$ is positive semidefinite, we obtain
\begin{multline}\label{eq:detLower}
 \det \dsdft{\Wb} \ge \det\left((2a_1-1)\dsdft{\Xb}+ \frac{1}{12m^2} I\right)\\
= \prod_{i=1}^L\left((2a_1-1)\mu_i(\theta) + \frac{1}{12m^2} \right).
\end{multline}


Now let $A$ be an $n\times n$ matrix and let $\Vert A\Vert \triangleq \sum_{i,j=1}^n |a_{i,j}|$ denote the \emph{$\ell_1$-matrix norm} of $A$. Since $\dsdft{\Nb}$ is positive semidefinite, the element with the maximum modulus is on the main diagonal; cf.~\cite[Problem~7.1.P1]{Horn_Matrix}. Furthermore, by assumption, on $\dom{F}_\Upsilon^\mathsf{c}$ the diagonal elements are bounded from above by $\frac{\Upsilon}{m^2}$. We hence obtain $\Vert \dsdft{\Nb} \Vert \le L^2\Upsilon/m^2\triangleq \eta_{\max}(\theta)$. It is known that all matrix norms\footnote{This bound holds without a multiplicative constant, since the spectral radius of a matrix is the infimum of all matrix norms~\cite[Lemma~5.6.10]{Horn_Matrix}.} bound the largest eigenvalue of the matrix from above~\cite[Th.~5.6.9]{Horn_Matrix}. Thus, $\eta_{\max}(\theta)$ is also an upper bound on the largest eigenvalue of $\dsdft{\Nb}$. Let $\omega_i(\theta)$, $i=1\dots,L$, denote the eigenvalues of $\dsdft{\Wb}$. Then we have for $m$ sufficiently large such that $2a_1-1\ge 0$~\cite[Cor.~4.3.15]{Horn_Matrix}

\begin{align}\label{eq:detUpper}
 \det \dsdft{\Wb} &= \prod_{i=1}^L \omega_i(\theta)\notag \\
&\le \prod_{i=1}^L \left( (2a_1-1)\mu_i(\theta) + \eta_{\max}(\theta) + \frac{1}{12m^2}\right)\notag \\
&\le \prod_{i=1}^L \left( (2a_1-1)\mu_i(\theta)  + \frac{\frac{1}{12}+L^2\Upsilon}{m^2}\right).
\end{align}
Combining~\eqref{eq:detLower} and~\eqref{eq:detUpper} with~\eqref{eq:integralToEvaluate}, we obtain
\begin{IEEEeqnarray}{lCl}
 \IEEEeqnarraymulticol{3}{l}{\limm \sum_{i=1}^L\frac{\int_{\dom{F}^c_\Upsilon} \log\left((2a_1-1)\mu_i(\theta) + \frac{1}{12m^2} \right) \diff\theta}{\log m}}\nonumber\\
 \quad &\le & \liminf_{m\to\infty} \frac{\int_{\dom{F}^c_\Upsilon} \log \det \dsdft{\Wb} \diff\theta}{\log m} \nonumber\\
 &\le & \limsup_{m\to\infty} \frac{\int_{\dom{F}^c_\Upsilon} \log \det \dsdft{\Wb} \diff\theta}{\log m} \nonumber\\
 &\le & \limm \sum_{i=1}^L\frac{\int_{\dom{F}^c_\Upsilon} \log\left((2a_1-1)\mu_i(\theta) + \frac{\frac{1}{12}+L^2\Upsilon}{m^2} \right) \diff\theta}{\log m}. \nonumber\\
\end{IEEEeqnarray}
It thus remains to evaluate 
 \begin{equation}\label{eq:integralReformulated}
L-\sum_{i=1}^L \limm \int_{\dom{F}^c_\Upsilon} \frac{\log \left((2a_1-1)\mu_i(\theta)  + \frac{\mathsf{K}}{m^2} \right)}{\log(1/m^2)} \diff\theta 
 \end{equation}
where $\mathsf{K}$ is either $1/12$ or $1/12+L^2\Upsilon$.

\subsection{Exchanging Limit and Integration}
To evaluate \eqref{eq:integralReformulated}, we continue along the lines of~\cite[Sec.~VIII]{lapidoth05}. Specifically, for each $i$, we split the integral on the RHS of~\eqref{eq:integralReformulated} into three parts: 
\begin{subequations}
 \begin{align}
  \dom{F}_I &\triangleq \{\theta\in\dom{F}_\Upsilon^\mathsf{c}{:}\ \mu_i(\theta)=0\}\\
\dom{F}_{II} &\triangleq \{\theta\in\dom{F}_\Upsilon^\mathsf{c}{:}\ \mu_i(\theta)\ge \mathsf{K}/(1-\varepsilon)\}\\
\dom{F}_{III} &\triangleq \{\theta\in\dom{F}_\Upsilon^\mathsf{c}{:}\ 0<\mu_i(\theta)<\mathsf{K}/(1-\varepsilon)\}
 \end{align}
\end{subequations}
where $0<\varepsilon\ll 1$ is arbitrary.

For the first part, we obtain
\begin{IEEEeqnarray}{lCl}
 \IEEEeqnarraymulticol{3}{l}{\int_{\dom{F}_I} \frac{\log \left((2a_1-1)\mu_i(\theta)  + \frac{\mathsf{K}}{m^2} \right)}{\log(1/m^2)} \diff\theta} \nonumber\\
\qquad & = & \int_{\dom{F}_I} \frac{\log \mathsf{K} + \log(1/m^2)}{\log(1/m^2)} \diff\theta \nonumber\\
& = & \lebesgue(\dom{F}_I)\left(1+\frac{\log \mathsf{K}}{\log(1/m^2)}\right)\label{eq:F1}
\end{IEEEeqnarray}
which evaluates to $\lebesgue(\dom{F}_I)$ in the limit as $m\to \infty$.

We next show that the integrals over $\dom{F}_{II}$ and $\dom{F}_{III}$ do not contribute to~\eqref{eq:integralReformulated}. To this end, it suffices to consider the integral of the function
\begin{equation}\label{eq:AmosFunction}
 \frac{\log \left((2a_1-1)\frac{\mu_i(\theta)}{\mathsf{K}}  + \frac{1}{m^2} \right)}{\log(1/m^2)} \triangleq \frac{\log \left(A_m(\theta)  + \frac{1}{m^2} \right)}{\log(1/m^2)}.
\end{equation}
In the remainder of the proof, we shall assume without loss of generality that $m^2> 8/\pi$, in which case $A_m(\theta)>0$ on $\theta\in\dom{F}_{II}\cup\dom{F}_{III}$. Clearly, whenever $A_m(\theta)>0$, the function in \eqref{eq:AmosFunction} converges to zero as $m\to \infty$. Moreover, for $A_m(\theta)\ge 1$, this function is nonpositive.

For all $\theta\in\dom{F}_{II}$ we have $A_m(\theta)\ge(2a_1-1)/(1-\varepsilon)$, hence we can find a sufficiently large $m_0$ such that, by \eqref{eq:boundBussgang} in Lemma~\ref{lem:quantizedProcess}, we have $A_m(\theta)\ge 1$, $m\geq m_0$. Since by the same result we also have $2a_1-1\le 2$, $m^2>8/\pi$, it follows that, for $m\geq m_0$,
\begin{multline}
 \frac{\log \left( 2\frac{\mu_i(\theta)}{\mathsf{K}}  + \frac{1}{m^2} \right)}{\log(1/m^2)}
\le \frac{\log \left(A_m(\theta)  + \frac{1}{m^2} \right)}{\log(1/m^2)} \le 0. \label{eq:this}
\end{multline}
The LHS of \eqref{eq:this} is nonpositive and monotonically increases to zero as $m\to \infty$.  We can thus apply the monotone convergence theorem to get
\begin{align}
 0 &\ge \limsup_{m\to\infty} \int_{\dom{F}_{II}}  \frac{\log \left(A_m(\theta)  + \frac{1}{m^2} \right)}{\log(1/m^2)}\diff\theta  \notag \\
& \ge \liminf_{m\to\infty} \int_{\dom{F}_{II}}  \frac{\log \left(A_m(\theta)  + \frac{1}{m^2} \right)}{\log(1/m^2)}\diff\theta  \notag \\
&\ge\limm \int_{\dom{F}_{II}}  \frac{\log \left(2\frac{\mu_i(\theta)}{\mathsf{K}}  + \frac{1}{m^2} \right)}{\log(1/m^2)}\diff\theta\notag\\
&=  \int_{\dom{F}_{II}} \limm \frac{\log \left( 2\frac{\mu_i(\theta)}{\mathsf{K}}  + \frac{1}{m^2} \right)}{\log(1/m^2)}\diff\theta \notag\\
&=0. \label{eq:F2}
\end{align}

We next turn to the case $\theta\in\dom{F}_{III}$. It was shown in~\cite[p.~443]{lapidoth05} that if $A_m(\theta)<1$, then the function in~\eqref{eq:AmosFunction} is bounded from above by 1. Furthermore, if $A_m(\theta)<1-\frac{1}{m^2}$ then it is nonnegative, and if $A_m(\theta)\geq 1-\frac{1}{m^2}$ then it is nonpositive and monotonically increasing in $m$. Restricting ourselves to the case $m^2>8/\pi$, we thus obtain for $\theta\in\dom{F}_{III}$
\begin{equation}
 \frac{\log \left(A_m(\theta)  + \frac{1}{m^2} \right)}{\log(1/m^2)} \ge 
 \begin{cases}
      \frac{\log \left(\frac{2}{1-\varepsilon}  + \frac{\pi}{8} \right)}{\log(\pi/8)}, & A_m(\theta)\ge 1-\frac{\pi}{8}\\
      0, & \text{otherwise}
 \end{cases}
\end{equation}
where we made use of the fact that $A_m(\theta)<(2a_1-1)/(1-\varepsilon)$, $\theta\in\dom{F}_{III}$ and, by \eqref{eq:boundBussgang} in Lemma~\ref{lem:quantizedProcess}, $2a_1-1\leq 2$, $m^2>8/\pi$. Hence, on $\dom{F}_{III}$ the magnitude of the function in \eqref{eq:AmosFunction} is bounded by
\begin{equation}\label{eq:AmosBound}
 \left|\frac{\log \left(A_m(\theta)  + \frac{1}{m^2} \right)}{\log(1/m^2)}\right| 
 \le \max\left\{1,\frac{\log \left(\frac{2}{1-\varepsilon}  + \frac{\pi}{8} \right)}{\log(8/\pi)}\right\}.
\end{equation}
We can thus apply the dominated convergence theorem to get
\begin{multline}\label{eq:F3}
  \limm \int_{\dom{F}_{III}}  \frac{\log \left((2a_1-1)\mu_i(\theta)  + \frac{\mathsf{K}}{m^2} \right)}{\log(1/m^2)}\diff\theta\\
 =  \int_{\dom{F}_{III}} \limm \frac{\log \left((2a_1-1)\frac{\mu_i(\theta)}{\mathsf{K}}  + \frac{1}{m^2} \right)}{\log(1/m^2)}\diff\theta 
 =0.
\end{multline}
Combining~\eqref{eq:F1},~\eqref{eq:F2}, and~\eqref{eq:F3}, we have that
\begin{IEEEeqnarray}{lCl}
\IEEEeqnarraymulticol{3}{l}{L+ \limm \frac{1}{2\log m}  \int_{\dom{F}^c_\Upsilon} \log \det \dsdft{\Wb} \diff\theta} \nonumber\\
\qquad & = & \sum_{i=1}^L \left( 1-\lebesgue(\{\theta\in\dom{F}_\Upsilon^\mathsf{c}{:}\ \mu_i(\theta)=0\})\right) \nonumber\\
& = & \sum_{i=1}^L \lebesgue(\{\theta\in\dom{F}_\Upsilon^\mathsf{c}{:}\ \mu_i(\theta)>0\}). \label{eq:after_lim}
\end{IEEEeqnarray}
By the continuity of the Lebesgue measure, this tends to $\sum_{i=1}^L \lebesgue(\{\theta{:}\ \mu_i(\theta)>0\})$ as $\Upsilon$ tends to infinity. But since the rank of a matrix is exactly the number of non-zero eigenvalues we obtain
\begin{equation}
 \sum_{i=1}^L \lebesgue(\{\theta{:}\ \mu_i(\theta)>0\})= \int_{-1/2}^{1/2} \mathrm{rank}(\dsdft{\Xb}) \diff \theta.
\end{equation}
To summarize, combining \eqref{eq:proof:Lplus}, \eqref{eq:throw_away}, \eqref{eq:F3}, and \eqref{eq:after_lim}, we obtain that
\begin{IEEEeqnarray}{lCl}
 d(\{\Xb_t\}) & = & L + \lim_{\Upsilon\to\infty}\limm \frac{1}{2\log m} \int_{-1/2}^{1/2} \log\det\dsdft{\Wb} \diff\theta \nonumber\\
 & = & \int_{-1/2}^{1/2} \mathrm{rank}(\dsdft{\Xb}) \diff \theta.
\end{IEEEeqnarray}
This completes the proof.

\section{Auxiliary Results}\label{app:aux}
In this section, we show that in the statement and proof of Theorem~\ref{thm:multivariate} we may assume without loss of generality that every component process has zero mean and unit variance. Moreover, we present the multivariate counterpart of~\cite[Lemma~6]{GeigerKoch_DimRate}.

\begin{lem}\label{lem:zerovariance}
 Suppose that  $\{\Xb_t\}$ is a stationary, $L$-variate Gaussian process with mean vector $\bfmu$ and SDF $\sdf{\Xb}$. Suppose that the component processes are ordered by their variances, i.e., $\infty>\sigma_1^2\ge \sigma_2^2\ge\cdots\sigma_{L'}^2\ge\sigma_{L'+1}^2=\cdots\sigma_L^2=0$ ($L'\le L$). Then, $\dimpinrate=d(\{\frac{1}{\sigma_1}X_{1,t},\dots,\frac{1}{\sigma_{L'}}X_{L',t}\})$ and, for almost every $\theta$, $\mathrm{rank}(\dsdft{\Xb})=\mathrm{rank}(\dsdft{(X_1/\sigma_1,\dots,X_{L'}/\sigma_{L'})})$.
\end{lem}

\begin{IEEEproof}
 Normalizing component processes with positive variance to unit variance does not affect the information dimension rate, as follows from Lemma~\ref{lem:scaleinv}. If $\sigma_i^2=0$, then the component process $\{X_{i,t}\}$ is almost surely constant. It follows that $\ent{\qRVm{X_{i,1}^k}}=0$ for every $m$ and every $k$, and hence, by the chain rule,
 \begin{multline}
 \ent{\qRVm{\Xb_1^k}} =  \ent{\qRVm{X_{1,1}^k},\dots,\qRVm{X_{L',1}^k}} \\+ \underbrace{\sum_{i=L'+1}^L \ent{\qRVm{X_{i,1}^k}|\qRVm{X_{1,1}^k},\dots,\qRVm{X_{i-1,1}^k}}}_{=0}.
 \end{multline}
 Dividing by $k\log m$ and letting $m$ and $k$ tend to zero shows that $\dimpinrate=d(\{\frac{1}{\sigma_1}X_{1,t},\dots,\frac{1}{\sigma_{L'}}X_{L',t}\})$.

Let $\Pi$ be an $L'\times L'$ diagonal matrix with values $\sigma_i$ on the main diagonal. For component processes with zero variance, the corresponding row and column of $\dsdft{\Xb}$ is zero almost everywhere. Hence, we have for almost every $\theta$ that
\begin{equation}
 \dsdft{\Xb} =\left[\begin{array}{cc}
               \Pi\dsdft{(X_1/\sigma_1,\dots,X_{L'}/\sigma_{L'})}\Pi & 0\\0 & 0
              \end{array}\right]
\end{equation}
where $0$ denotes an all-zero matrix of appropriate size. We thus have $\mathrm{rank}(\dsdft{\Xb})=\mathrm{rank}(\dsdft{(X_1/\sigma_1,\dots,X_{L'}/\sigma_{L'})})$ for almost every $\theta$.
\end{IEEEproof}

\begin{lem}\label{lem:KLDBounded}
	Let $\Xb$ be an $\ell$-variate Gaussian vector with mean vector $\bfmu$ and covariance matrix $\covmat{\Xb}$. Let $\Wb\triangleq[\Xb]_m+\Ub$, where $\Ub$ is a $\ell$-variate vector, independent of $\Xb$, with components independently and uniformly distributed on $[0,1/m)$.
    Then,
	\begin{equation}
	\frac{\kld{\pdf{\Wb}}{\pdfg{\Wb}}}{\ell} \le \frac{1}{2}\log\left(2 \pi \left(1+\frac{1}{12}\right)\right) + \frac{75}{2} + \frac{24}{\pi}.
	\end{equation}
\end{lem}

\begin{IEEEproof}
 By \cite[Th.~23.6.14]{Lapidoth_DigitalCommunication}, the vector $\Xb=\trans{(X_1,\ldots,X_\ell)}$ can be written as
\begin{equation}
\label{eq:XbNb}
\Xb = A \Nb + \bfmu
\end{equation}
where $\Nb$ is a $\ell'$-dimensional, zero-mean, Gaussian vector ($\ell'\leq \ell$) with independent components whose variances are the nonzero eigenvalues of $\covmat{\Xb}$ and where $A$ is a $\ell\times \ell'$ matrix satisfying $\trans{A} A=I$ (with $I$ denoting the identity matrix). We use the data processing inequality, the chain rule for relative entropy, and the fact that $\Nb$ is Gaussian to obtain
\begin{IEEEeqnarray}{lCl}
 \kld{\pdf{\Wb}}{\pdfg{\Wb}} & \le & \kld{\pdf{\Wb,\Nb}}{\pdfg{\Wb,\Nb}}\nonumber\\
& \le & \kld{\pdf{\Nb}}{\pdfg{\Nb}} \nonumber\\
& & {} + \int \kld{\pdf{\Wb|\Nb=\mathbf{n}}}{\pdfg{\Wb|\Nb=\mathbf{n}}} \pdf{\Nb}(\mathbf{n}) \diff \mathbf{n} \nonumber\\
& = & \int \kld{\pdf{\Wb|\Nb=\mathbf{n}}}{\pdfg{\Wb|\Nb=\mathbf{n}}} \pdf{\Nb}(\mathbf{n}) \diff \mathbf{n}\label{eq:klddpi}
\end{IEEEeqnarray}
where $\pdfg{\Wb,\Nb}$ denotes the PDF of a Gaussian vector with the same mean vector and covariance matrix as $(\Wb,\Nb)$, and
\begin{subequations}
\begin{align}
\pdf{\Wb|\Nb=\mathbf{n}}(\mathbf{w}) &\triangleq \frac{\pdf{\Wb,\Nb}(\mathbf{w},\mathbf{n})}{\pdfg{\Nb}(\mathbf{n})}\\
 \pdfg{\Wb|\Nb=\mathbf{n}}(\mathbf{w}) &\triangleq \frac{\pdfg{\Wb,\Nb}(\mathbf{w},\mathbf{n})}{\pdfg{\Nb}(\mathbf{n})}.
\end{align}
\end{subequations}

To evaluate the relative entropy on the RHS of \eqref{eq:klddpi}, we first note that, given $\Xb$, the random vector $\Wb$ is uniformly distributed on an $\ell$-dimensional cube of length $\frac{1}{m}$. Since $\Xb$ can be obtained from $\Nb$ via \eqref{eq:XbNb}, the conditional PDF of $\Wb$ given $\Nb=\mathbf{n}$ is
\begin{equation}
\pdf{\Wb|\Nb=\mathbf{n}}(\mathbf{w}) =  m^\ell \mathbf{1}\{\qRV{\mathbf{w}}{m} = \qRV{A\mathbf{n}+\bfmu}{m}\}, \quad \mathbf{w} \in\mathbb{R}^\ell.
\end{equation}
Consequently, denoting $\mathbf{z}=\qRV{A\mathbf{n}+\bfmu}{m}$,
\begin{multline}
\kld{\pdf{\Wb|\Nb=\mathbf{n}}}{\pdfg{\Wb|\Nb=\mathbf{n}}}  = \log\left( m^\ell \sqrt{(2\pi)^\ell\det\covmat{\Wb|\Nb} }\right)\\
 + \frac{m^\ell}{2} \int_{\dom{C}(\mathbf{z},1/m)} \trans{(\mathbf{w}-\bfmu_{\Wb|\Nb=\mathbf{n}})}\covmat{\Wb|\Nb}^{-1}(\mathbf{w}-\boldsymbol{\mu}_{\Wb|\Nb=\mathbf{n}})  \diff \mathbf{w}  \label{eq:kld_cond}
\end{multline}
where $\bfmu_{\Wb|\Nb=\mathbf{n}}$ and $\covmat{\Wb|\Nb}$ denote the conditional mean and the conditional covariance matrix of $\Wb$ given $\Nb=\mathbf{n}$. These can be computed as \cite[Th.~23.7.4]{Lapidoth_DigitalCommunication}
\begin{subequations}
\label{eq:WbNb}
\begin{align}
 \bfmu_{\Wb|\Nb=\mathbf{n}} & = \expec{\Wb} + \covmat{\Wb \Nb}\covmat{\Nb}^{-1} \mathbf{n} \label{eq:mu_WbNb} \\
 \covmat{\Wb|\Nb} & = \covmat{\Wb} - \covmat{\Wb \Nb}\covmat{\Nb}^{-1}\trans{\covmat{\Wb \Nb}} \label{eq:C_WbNb}
\end{align}
\end{subequations}
where $\covmat{\Wb \Nb}$ denotes the cross-covariance matrix of $\Wb$ and $\Nb$, and $\covmat{\Wb}$ and $\covmat{\Nb}$ denote the covariance matrices of $\Wb$ and $\Nb$, respectively.

Letting $\Zb=[\Xb]_m$, we have $\Wb=\Zb+\Ub$. Since $\Ub$ is independent of $\Xb$, the cross-covariance matrix of $\Wb$ and $\Xb$ (denoted by $\covmat{\Wb\Xb}$) is equal to the cross-covariance matrix of $\Zb$ and $\Xb$ (denoted by $\covmat{\Zb\Xb}$). Bussgang's theorem~\cite[eq.~(19)]{bussgang52} yields $\covt{Z_jX_i} = a_{j}  \covt{X_jX_i}$ where $a_j$ is as in Lemma~\ref{lem:quantizedProcess}. Hence, if $\Lambda_\mathbf{a}$ is a diagonal matrix with the elements of $\mathbf{a}=(a_1,\dots,a_\ell)$ on the main diagonal, then $\covmat{\Zb\Xb}=\Lambda_\mathbf{a} \covmat{\Xb}$.  From~\eqref{eq:XbNb} we get $\covmat{\Xb}=A\covmat{\Nb}\trans{A}$ and $\covmat{\Wb\Nb}=\covmat{\Wb\Xb}A$, hence
\begin{equation}
 \covmat{\Wb\Nb}=\covmat{\Wb\Xb}A=\covmat{\Zb\Xb}A=\Lambda_\mathbf{a} \covmat{\Xb}A=\Lambda_\mathbf{a}A\covmat{\Nb}.
\end{equation}
Together with \eqref{eq:WbNb}, this yields
\begin{subequations}
\begin{IEEEeqnarray}{rCl}
 \bfmu_{\Wb|\Nb=\mathbf{n}} & = & \expec{\Wb} + \Lambda_\mathbf{a} A \mathbf{n} \label{eq:mu_WbNb_2}\\
 \covmat{\Wb|\Nb} & = & \covmat{\Wb} - \Lambda_\mathbf{a}\covmat{\Xb}\Lambda_\mathbf{a}. \label{eq:cov_wbnb_2}
\end{IEEEeqnarray}
\end{subequations}

Combining \eqref{eq:mu_WbNb_2}  with \eqref{eq:XbNb}, and using the triangle inequality, we upper-bound each component of $\mathbf{w}-\bfmu_{\Wb|\Nb=\mathbf{n}}$ as
\begin{multline}
\left|w_{j} - \expec{W_{j}} - a_j(x_{j}-\mu_j)\right|   \leq  |z_{j}-x_{j}| + \left|u_{j} - \expec{U_{j}}\right|\\
+ \left|\expec{Z_{j}}-\mu_j\right| + |1-a_j| |x_{j}-\mu_j|. \label{eq:difference_ISIT}
\end{multline}
The first and the third term on the RHS of \eqref{eq:difference_ISIT} are both upper-bounded by $\frac{1}{m}$, and the second term is upper-bounded by $\frac{1}{2m}$. From \eqref{eq:boundBussgang} in Lemma~\ref{lem:quantizedProcess} we get that the term $|1-a_j|$ is upper bounded by $1/m\sqrt{2/\pi\sigma_j^2}$, where $\sigma_j^2$ is the variance of $X_j$. We thus obtain
\begin{equation}
\label{eq:bound_norm}
\|\mathbf{w}-\bfmu_{\Wb|\Nb=\mathbf{n}}\|^2 \leq \frac{1}{m^2}\left(\frac{25k}{2} + \frac{4}{\pi}\sum_{j=1}^\ell \frac{(x_j-\mu_j)^2}{\sigma_j^2} \right).
\end{equation}
We next note that, since $\Wb=\Zb+\Ub$ and since $\Ub$ is independent from $\Zb$ and i.i.d. on $[0,1/m)$,
\begin{equation}
\label{eq:covmat_lemma1}
\covmat{\Wb|\Nb} = \covmat{\Zb} - \Lambda_\mathbf{a}\covmat{\Xb}\Lambda_\mathbf{a} + \frac{1}{12m^2} I.
\end{equation}
It can be shown that $\covmat{\Zb} - \Lambda_\mathbf{a} \covmat{\Xb}\Lambda_\mathbf{a}$ is the conditional covariance matrix of $\Zb$ given $\Nb$, hence it is positive semidefinite.\footnote{Indeed, we have $\covmat{\Zb\Xb}=\covmat{\Wb\Xb}$ and, by \eqref{eq:XbNb}, $\covmat{\Zb\Nb}=\covmat{\Zb\Xb}A$. Replacing in \eqref{eq:C_WbNb} $\Wb$ by $\Zb$, and repeating the steps leading to \eqref{eq:cov_wbnb_2}, we obtain the desired result.} It follows that the smallest eigenvalue of $\covmat{\Wb|\Nb}$ is lower-bounded by $\frac{1}{12 m^2}$. Together with \eqref{eq:bound_norm}, this yields for the second term on the RHS of \eqref{eq:kld_cond}
\begin{IEEEeqnarray}{lCl}
\IEEEeqnarraymulticol{3}{l}{\frac{m^\ell}{2} \int_{\dom{C}(\mathbf{z},1/m)} \trans{(\mathbf{w}-\bfmu_{\Wb|\Nb=\mathbf{n}})}\covmat{\Wb|\Nb}^{-1}(\mathbf{w}-\bfmu_{\Wb|\Nb=\mathbf{n}})  \diff \mathbf{w}}\nonumber \\
\qquad & \leq &  6 m^{\ell+2} \frac{1}{m^\ell} \frac{1}{m^2}\left(\frac{25\ell}{2} + \frac{4}{\pi}\sum_{j=1}^\ell \frac{(x_j-\mu_j)^2}{\sigma_j^2}\right)\nonumber\\
 & = & \frac{75\ell}{2} + \frac{24}{\pi}\sum_{j=1}^\ell \frac{(x_j-\mu_j)^2}{\sigma_j^2}. \label{eq:split_1}
\end{IEEEeqnarray}

To upper-bound the first term on the RHS of \eqref{eq:kld_cond}, we use that \eqref{eq:covmat_lemma1} combined with Lemma~\ref{lem:quantizedProcess} implies that every diagonal element of $\covmat{\Wb|\Nb}$ is given by
\begin{multline}
\expec{(Z_j-\expec{Z_j})^2} - a_j^2\sigma_j^2 + \frac{1}{12m^2}
 =  -(1-a_j)^2 \sigma_j^2 \\ {} + \expec{(X_j-\mu_j-Z_j+\expec{Z_j})^2} + \frac{1}{12m^2}. \label{eq:dimrateproof:moreslowly}
\end{multline}
The first term on the RHS of \eqref{eq:dimrateproof:moreslowly} is negative and the second term is upper-bounded by $ \expec{(X_j-Z_j)^2}\le 1/m^2$. Hence, every element on the main diagonal of $\covmat{\Wb|\Nb}$ is upper-bounded by $\frac{1+1/12}{m^2}$. It thus follows from Hadamard's inequality that
\begin{equation}
\label{eq:split_2}
\log\left( m^\ell \sqrt{(2\pi)^\ell\det\covmat{\Wb|\Nb} }\right) \leq \frac{\ell}{2} \log\left(2 \pi \left(1+\frac{1}{12}\right)\right).
\end{equation}
Combining \eqref{eq:split_1} and \eqref{eq:split_2} with \eqref{eq:kld_cond} and \eqref{eq:klddpi} yields
\begin{IEEEeqnarray}{lCl}
\IEEEeqnarraymulticol{3}{l}{\kld{\pdf{\Wb_1^\ell}}{\pdfg{\Wb_1^\ell}}}\nonumber\\
\,\,\, & \leq & \frac{\ell}{2} \log\left(2 \pi \left(1+\frac{1}{12}\right)\right) + \frac{75\ell}{2} + \frac{24}{\pi}\sum_{j=1}^\ell \frac{\expec{(X_j-\mu_j)^2}}{\sigma_j^2}\nonumber\\
& = & \ell\left(\frac{1}{2}\log\left(2 \pi \left(1+\frac{1}{12}\right)\right) + \frac{75}{2} + \frac{24}{\pi} \right).
\end{IEEEeqnarray}
This completes the proof.
\end{IEEEproof}

\bibliographystyle{IEEEtran}
\bibliography{IEEEabrv,references}

\begin{thebibliography}{10}
\providecommand{\url}[1]{#1}
\csname url@samestyle\endcsname
\providecommand{\newblock}{\relax}
\providecommand{\bibinfo}[2]{#2}
\providecommand{\BIBentrySTDinterwordspacing}{\spaceskip=0pt\relax}
\providecommand{\BIBentryALTinterwordstretchfactor}{4}
\providecommand{\BIBentryALTinterwordspacing}{\spaceskip=\fontdimen2\font plus
\BIBentryALTinterwordstretchfactor\fontdimen3\font minus
  \fontdimen4\font\relax}
\providecommand{\BIBforeignlanguage}[2]{{%
\expandafter\ifx\csname l@#1\endcsname\relax
\typeout{** WARNING: IEEEtran.bst: No hyphenation pattern has been}%
\typeout{** loaded for the language `#1'. Using the pattern for}%
\typeout{** the default language instead.}%
\else
\language=\csname l@#1\endcsname
\fi
#2}}
\providecommand{\BIBdecl}{\relax}
\BIBdecl

\bibitem{Renyi_InfoDim}
A.~R\'{e}nyi, ``On the dimension and entropy of probability distributions,''
  \emph{Acta Mathematica Hungarica}, vol.~10, no. 1-2, pp. 193--215, Mar. 1959.

\bibitem{Kawabata_RDDim}
T.~Kawabata and A.~Dembo, ``The rate-distortion dimension of sets and
  measures,'' \emph{{IEEE} Trans. Inf. Theory}, vol.~40, no.~5, pp. 1564--1572,
  Sep. 1994.

\bibitem{Koch_SLB}
T.~Koch, ``The {Shannon} lower bound is asymptotically tight,'' \emph{{IEEE}
  Trans. Inf. Theory}, vol.~62, no.~11, pp. 6155--6161, Nov. 2016.

\bibitem{Wu_Renyi}
Y.~Wu and S.~Verd\'{u}, ``R\'{e}nyi information dimension: Fundamental limits
  of almost lossless analog compression,'' \emph{{IEEE} Trans. Inf. Theory},
  vol.~56, no.~8, pp. 3721--3748, Aug. 2010.

\bibitem{Wu_IF}
Y.~Wu, S.~Shamai~(Shitz), and S.~Verd\'{u}, ``Information dimension and the
  degrees of freedom of the interference channel,'' \emph{{IEEE} Trans. Inf.
  Theory}, vol.~61, no.~1, pp. 256--279, Jan. 2015.

\bibitem{Stotz_IF}
D.~Stotz and H.~B\"olcskei, ``Degrees of freedom in vector interference
  channels,'' \emph{{IEEE} Trans. Inf. Theory}, vol.~62, no.~7, pp. 4172--4197,
  Jul. 2016.

\bibitem{GeigerKoch_ISIT}
B.~C. Geiger and T.~Koch, ``On the information dimension rate of stochastic
  processes,'' in \emph{Proc. IEEE Int. Symp. Inf. Theory (ISIT)}, Aachen,
  Germany, Jun. 2017, pp. 888--892.

\bibitem{GeigerKoch_DimRate}
------, ``On the information dimension of stochastic processes,'' {\tt
  arXiv:1702.00645v2 [cs.IT]}, Feb. 2017.

\bibitem{Wu_PhD}
Y.~Wu, ``Shannon theory for compressed sensing,'' Ph.D. dissertation, Princeton
  University, 2011.

\bibitem{Cover_Information1}
T.~M. Cover and J.~A. Thomas, \emph{Elements of Information Theory},
  1st~ed.\hskip 1em plus 0.5em minus 0.4em\relax Wiley Interscience, 1991.

\bibitem{gallager68}
R.~G. Gallager, \emph{Information Theory and Reliable Communication}.\hskip 1em
  plus 0.5em minus 0.4em\relax John Wiley \& Sons, 1968.

\bibitem{Wiener_StochasticProcesses}
N.~Wiener and P.~Masani, ``The prediction theory of multivariate stochastic
  processes,'' \emph{Acta Mathematica}, vol.~98, no.~1, pp. 111--150, 1957.

\bibitem{bussgang52}
J.~J. Bussgang, ``Crosscorrelation functions of amplitude-distorted {Gaussian}
  signals,'' Res.\ Lab.\ of Electronics, M.I.T., Cambridge, Mass., Tech. Rep.
  216, Mar. 1952.

\bibitem{Lapidoth_DigitalCommunication}
A.~Lapidoth, \emph{A Foundation in Digital Communication}.\hskip 1em plus 0.5em
  minus 0.4em\relax Cambridge: Cambridge University Press, 2009.

\bibitem{Neeser_Proper}
F.~D. Neeser and J.~L. Massey, ``Proper complex random processes with
  applications to information theory,'' \emph{{IEEE} Trans. Inf. Theory},
  vol.~39, no.~4, pp. 1293--1302, Jul. 1993.

\bibitem{Horn_Matrix}
R.~A. Horn and C.~R. Johnson, \emph{Matrix Analysis}, 2nd~ed.\hskip 1em plus
  0.5em minus 0.4em\relax Cambridge: Cambridge University Press, 2013.

\bibitem{lapidoth05}
A.~Lapidoth, ``On the asymptotic capacity of stationary {G}aussian fading
  channels,'' \emph{{IEEE} Trans. Inf. Theory}, vol.~51, no.~2, pp. 437--446,
  Feb. 2005.

\end{thebibliography}

\end{document}